# Development of the $\gamma$ strength function with the neutron number

*Stefan* Frauendorf[1], *Ronald* Schwengner[2], *Gowhar* Bhat[34,] and *Javid* Sheik[34]

[1]Department of Physics and Astronomy , University of Notre Dame, Notre Dame, Indiana 46556, USA
[2]Helmholtz Zentrum Dresden Rossendorf, 01328 Dresden, Germany
3Department of Physics, SP College Srinagar, Jammu and Kashmir, 190 001, India
[4]Cluster University Srinagar, Jammu and Kashmir, Srinagar, Goji Bagh, 190 008, India

**Abstract.** The M1 and E2 $\gamma$ strength functions ($\gamma sf$) have been calculated for extended series of the Mo, Fe, Sn, Ge and Gd isotopes using the conventional spherical shell model (SSM) and, as a new tool, the triaxial projected shell model (TPSM). For almost all cases the strong enhancement of the M1 $\gamma sf$ (low energy magnetic radiation-LEMAR) is found. In the mid-shell region, a bimodal structure of the LEMAR spike and a bump around 3 MeV, interpreted as the scissors resonance (SR), develops. The combination of LEMAR and the SR is generated by the splitting of the spherical single particle multiplets of given j caused by deformation and their fragmentation over nearby quasiparticle configurations.

## 1 Introduction

The $\gamma$ strength function ($\gamma sf$) measures the sum of the reduced transition probabilities of given multipolarity from an initial state $i$ to all states $\lambda$ within a narrow energy interval centred around $E_\lambda$ divided by the interval's width

$$f_1(E_\gamma) = \frac{\bar{\Gamma}}{E_\gamma^3}\rho(E_\lambda) = \overline{B(M1)}\rho(E_\lambda),\ E_\gamma = E_\lambda - E_i \quad (1)$$

where $\bar{\Gamma}$ is the radiation width averaged over the interval's states [1]. Usually, the M1 γsf is given in the units MeV$^{-3}$ based on the unit MeV$^{-2}$ for the average reduced width, which introduces the conversion factor of 1.15 10$^{-8}$ compared to using the common unit $\mu^2$ for $\overline{B(M1)}$. The black arrows in Fig. 1 display $\vec{f}_1(E_\gamma)$ [1] for the common case $E_i = 0$ of absorption of $\gamma$ rays from the ground state, where $\vec{f}_1$ is proportional to the absorption cross section $\sigma_\gamma$.

For astrophysical and technical applications, the $\gamma sf$ for the decay from excited states, $\overleftarrow{f}_1(E_i)$ [1], plays the deceive role. It is the sum of the reduced transition probabilities of given multipolarity from all states within a narrow energy interval around $E_\lambda$ to decay to an excited state at $E_i$, divided by the interval's width. The $E_\lambda$ interval represents the compound nucleus' energy and $\overleftarrow{f}_1(E_\gamma)$ is $\overleftarrow{f}_1(E_i)$ taken at $E_i = E_\lambda - E_\gamma$.

As illustrated by Fig. 1, the compound nucleus deexcites via $\gamma$ cascades, which start with different primary transitions. The average of the $\gamma sf$ for these transitions and the density of the populated states determine whether the nucleus deexcites via $\gamma$ emission or the reaction takes another path.

------------------

* Corresponding author: sfrauend@nd.edu

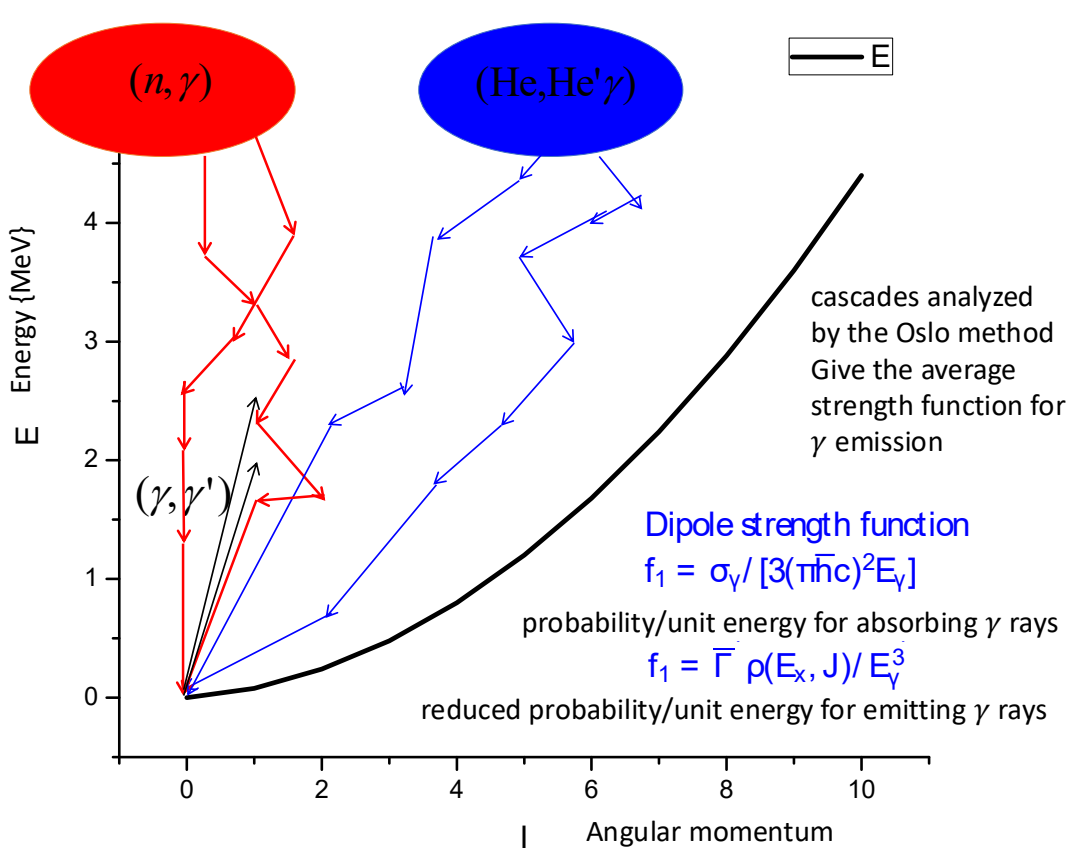

**Fig. 1.** The $\gamma$ absorption and emission processes.

The absorption γsf has been extensively studied. It is dominated by the Isovector Giant Dipole Resonance (IVGDR) around 15 MeV, which extends into the region $E_\gamma$ <8 MeV, below the neutron emission threshold. There, additional E1 strength has been identified, which has been associated with oscillations of the nuclear surface relative to its interior and octupole correlations. M1 strength has been found around 3 MeV, called Scissors Resonance (SR) and interpreted as orientation oscillations of the deformed neutron system against the deformed proton system, and around 8 MeV, called Spin Flip Resonance (SFR) and interpreted as oscillations of the nucleons' spins relative to each other.

Studies of the emission γsf found evidence for these phenomena as well. In addition, the Oslo collaboration observed an enhancement of the emission strength below 2 MeV, which has been called Low Energy Enhancement (LEE) [2] later. Fig. 2 shows an example.

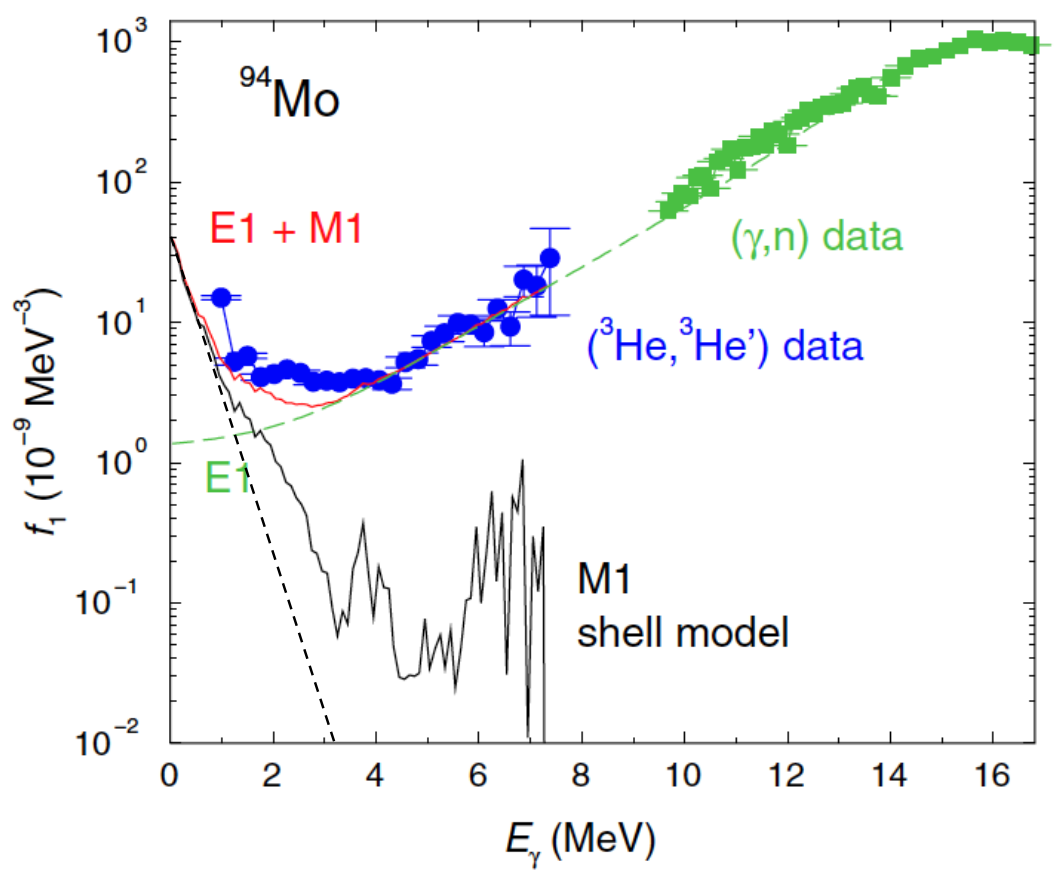

**Fig. 2.** The emission $\gamma sf$ for $^{94}$Mo from Ref. [2] (blue circles) compared with the SSM calculations of Ref. [3] (black curve) and the fit by Eq. (2) (dashed line). The green curve shows a General Lorentzian fit to the IVGR data (green circles) and the red curve the sum of the E1 and M1 contributions. Adapted from Ref. [3].

The observation came as a surprise because it seemed to be in conflict with the Brink-Axel hypothesis, which states that the $\gamma sf$ of excited states should not depend on the energy of the excited state.

## 2 The Mo and Fe isotopes

The authors of Ref. [3] firstly demonstrated that the M1 emission $\gamma sf$ showed the LEE indeed. By means of the Spherical Shell Model (SSM) they calculated the energies of the 40 lowest states of angular momentum $J = 0, \ldots, 6$ in a suitable valence space and the matrix of $B\{M1, \lambda \rightarrow i\}$ of the transitions between them (~15000). A matrix $[\lambda, i]$ was created by sorting the data in energy bins of 100 keV from which it was straightforward to calculate the $\gamma sf$ averaged over the $J$ and $i$ values.

As seen in Fig. 1, there is a spike at $E_\gamma = 0$, which in the region below 1 MeV can be approximated by

$$\overleftarrow{f}_1(E_\gamma) = f_0 e^{-E_\gamma/T_B} \qquad (2)$$

with $f_0 = 4\ 10^{-8}\text{MeV}^{-3}$ and $T_B$ =0.38. The exponential fall-off points to a stochastic mechanism that is characterized by $T_B$. The study explained the appearance of the LEE for the first time as magnetic radiation and the authors suggested the more specific name Low Energy Magnetic Radiation (LEMAR) for the LEE. The studied Mo isotopes with N=52, 53 and 54 are spherical and only the LEMAR spike appeared in the $\gamma sf$.

In order to explore the consequences of deformation, the authors of Ref. [4] firstly explored the open neutron shell in the Fe isotopes by means of the SSM using the same method as in Ref. [3] but another suitable valence space and Hamiltonian.

Fig. 2 summarizes the results. The near-spherical isotope $^{60}Fe_{34}$ has a single LEMAR spike that can be approximated by the expression (2) with $f_0 = 7 \times 10^{-8}$ MeV$^{-3}$ and $T_B$ =1.2 MeV up to 2 MeV. The deformed isotope $^{68}Fe_{42}$ shows the expected SM at 3 MeV **and** a weaker LEMAR spike., which can be approximated up to 1 MeV by Eq. (2) with $f_0 = 3.8 \times 10^{-8}$ MeV and $T_B$ =1.2 MeV. The M1 strength integrated from 0 to 2 MeV, $B(M1)_{\text{LEMAR}}$, and the strength integrated from 2 to 5 MeV, $B(M1)_{\text{SR}}$, are, respectively, 5.7 $\mu^2$and 3.5 $\mu^2$ for $^{60}Fe_{34}$ and 4.0 $\mu^2$and 6.5 $\mu^2$ for $^{68}Fe_{42}$. Their sums change weakly with $N$.

The figure also shows the E2 emission $\gamma sf$ $\overleftarrow{f}_2(E_\gamma)E_\gamma^2$. (The energy factor was introduced to have the same scale as $\overleftarrow{f}_1(E_\gamma)$.) The E2 bump seen for $N$=42 is caused by enhanced stretched E2 transitions, which signal the presence of deformation. The bump is absent for $N$=34. The absorption spectrum shows only lines above 2 MeV. Their $B(M1)$ sums up to $E_\gamma = 5$ MeV amount to 0.7 $\mu^2$ and 1.7 $\mu^2$ for $N$=34 and 42, respectively, which is much smaller than the integrated absorption strengths within this interval.

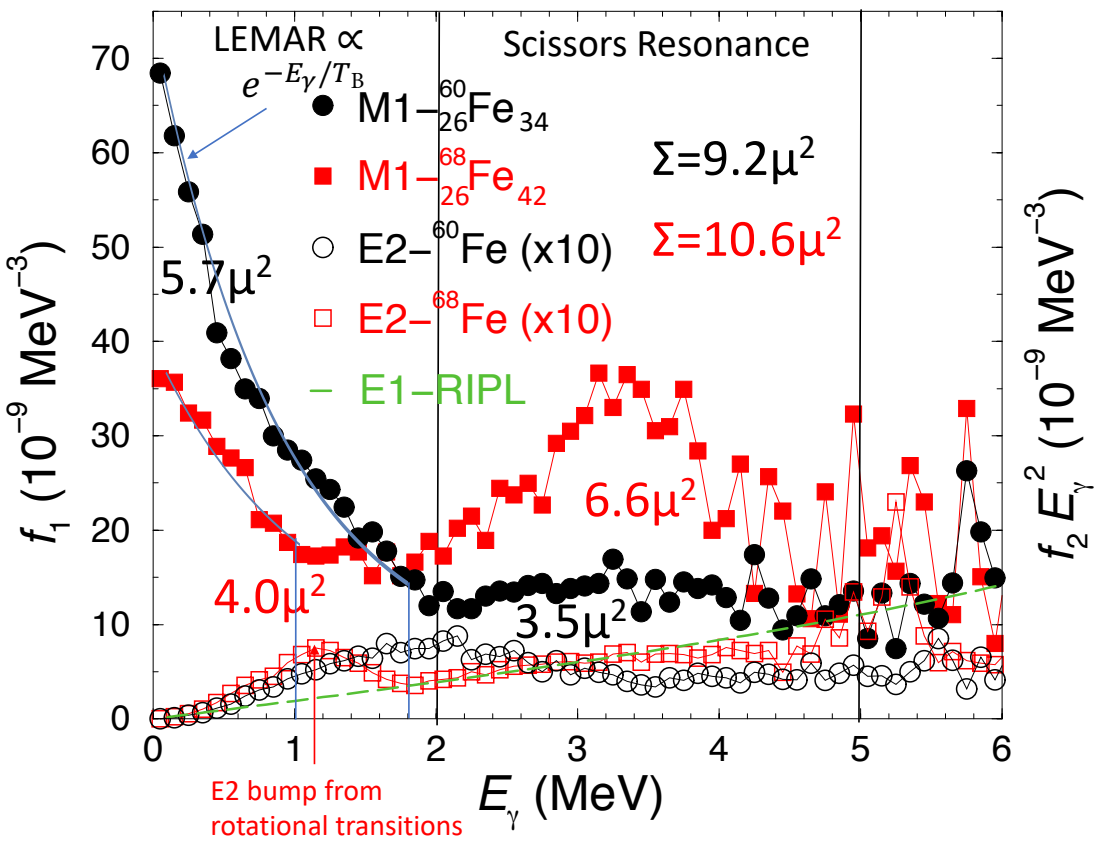

**Fig. 3.** The emission $\gamma sf$ for $^{60,68}$Fe calculated by means of the SSM in Ref. [4]. Black symbols depict the result for $N$=34 and red symbols for $N$=42. The thin curves show the fit by Eq. (2). The open symbols display the E2 $\gamma sf$ $\overleftarrow{f}_2(E_\gamma)E_\gamma^2$. Adapted from Ref. [4].

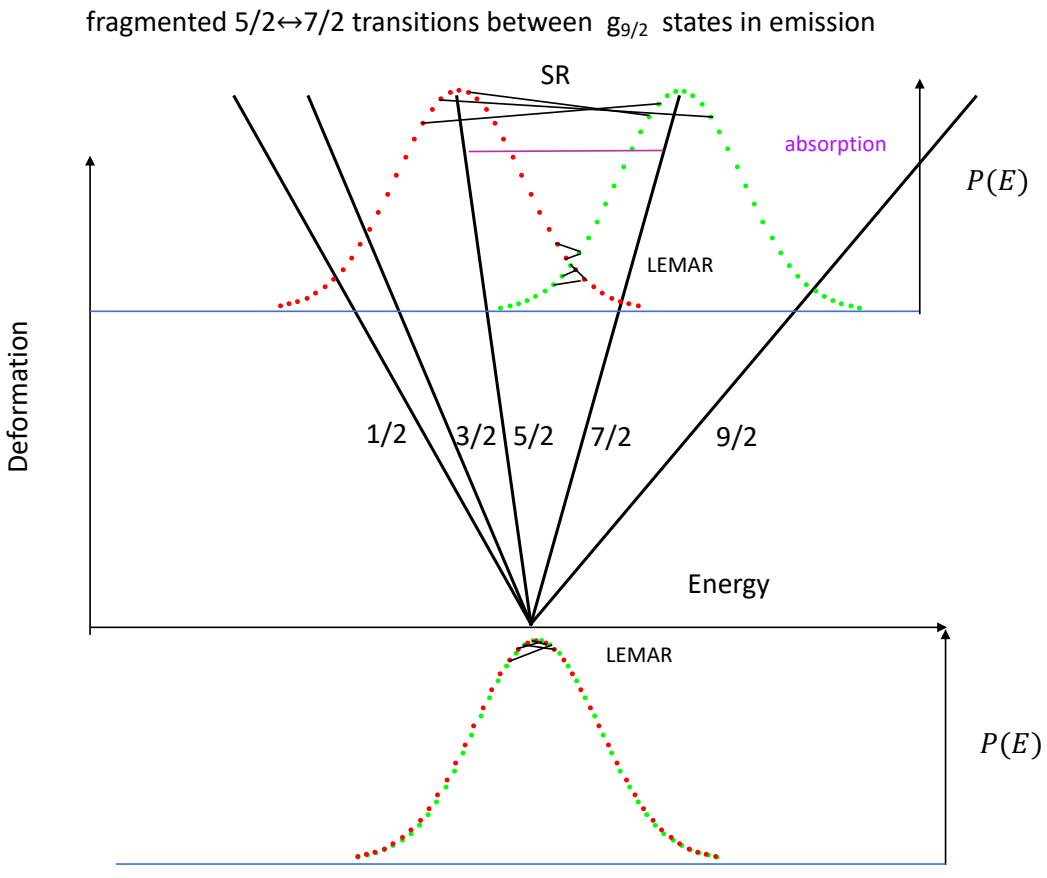

**Fig. 4.** Schematic illustration of the splitting by deformation (thick black lines) and fragmentation of a j-shell single particle state. The red dots display the Breit-Wigner distribution of the *k=5/2* single particle component over the complex mixed states and the red dots display the distribution of the *k*=7/2 single particle component, where *P(E)* is their probability. The thin black lines show some of the possible transitions between the complex states. The magenta line represents one of the two-quasiparticle lines of the SR in the absorption spectrum.

The Mo and Fe examples illustrate the general $N$ dependence of the emission $\gamma sf$. Fig. 4 explains the mechanism behind. It depicts how the magnetic substates $k$ of a spherical $g_{9/2}$ shell split with deformation, which is a schematic representation of a Nilsson diagram.

The SR in the absorption spectrum is composed of two-quasiparticle excitations corresponding to the $\Delta k = \pm 1$ particle-hole excitations in the figure. The magenta line indicates one of them. The two-quasiparticle energy is larger than two times the pairing gap and approaches the particle-hole energy with the excitation energy. Thus, the SR is expected at some average of the two-quasiparticle energies. The pair correlations reduce the $B(M1)$ of the two-quasiparticle transitions as compared to the particle-hole transitions ("uv" factors). The residual interaction, which is thought to be repulsive, may shift the SR lines somewhat up. However, the SSM calculations and the observed line spectra do not provide evidence for the existence of one highly collective scissors state that collects most of the M1 strength.

Most of the transitions that generate the emission $\gamma sf$ connect highly excited states, for which the pair correlations are quenched. As discussed in Ref. [5], the residual interaction causes a substantial fragmentation of the particle-hole strength over the neighbouring "background" configurations, which is given by a Breit-Wigner distribution.

Consider a $k$=5/2 and a $k$=7/2 particle added to the same configuration of the remaining $N$-1 nucleons. The $B(M1)$ value between the two states is equal to the single particle value. The coupling of these states to the other configurations of the $N$ nucleons will distribute the selected two configurations over the dense background states. The resulting final and initial states, which are displayed by the green and red dots in the figure, contain a small component of the selected states with the amplitudes $c_{5/2}$ and $c_{7/2}$, which contribute the fraction $|c_{5/2}c_{7/2}|B(M1)$ to the transitions between the complex states. Some of them are included in Fig. 4 where their length indicates the transition energy and the elevation the probability of admixture.

The lower part of Fig. 4 illustrates the case of no deformation. The single particle states $k$ are degenerated. The fragmentation distributes the $k$=5/2 and $k$=7/2 configurations over the mixed states, which have different energies. Low-energy transitions between them appear, which are displayed by the short lines between the red and green points. They represent the single LEMAR spike in spherical nuclei.

The upper part of Fig. 4 illustrates the case of large deformation. The long lines show the fragmented transitions of the $k$=5/2 and $k$=7/2 contributions to the SR, which loses its discrete line structure. In addition, there is the group of transitions around $E_\gamma$=0 between the overlapping distributions, which form the LEMAR spike that coexists with the SR.

The exponential fall-off of the LEMAR reflects the complex nature of the states between the transitions occur. They are localized in a region around 5 MeV (see e.g. Fig. 3 of Ref. [4]), As discussed in the review [6], their structure becomes chaotic at these energies. The amplitudes $c_{\frac{5}{2}}$ and $c_{\frac{7}{2}}$ fluctuate around 0 with random signs and $|c_k| \sim \frac{1}{\sqrt{D}}$, where $D$ is the dimension of the configuration space. Average quantities of two-body operators, like $B(M1)$, between nearby states can be evaluated as expectation values of a statistical ensemble that is characterised by a chaoticity parameter $T_B$

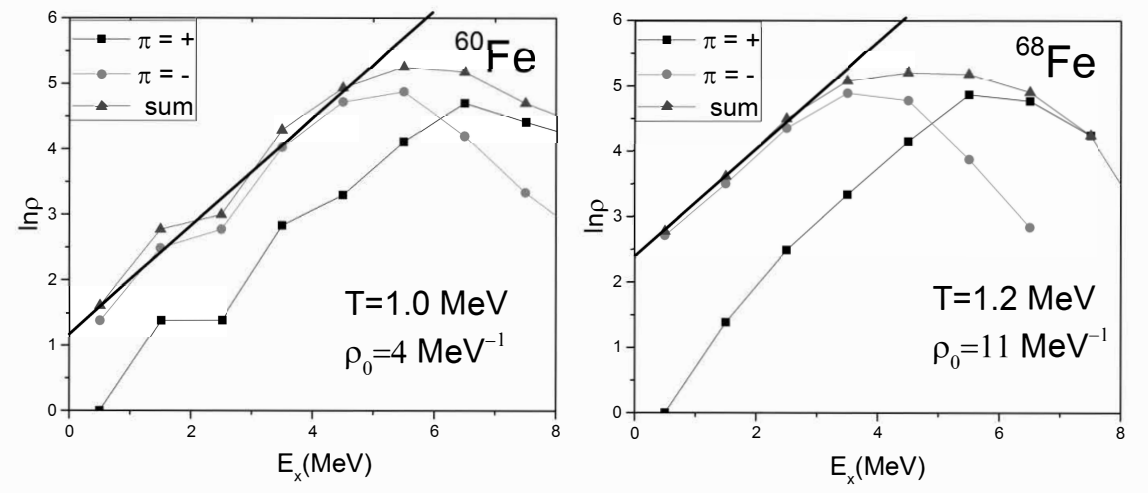

**Fig. 5.** Level density of [60,68]Fe calculated by means of the SSM in Ref. [4]. The straight line shows the constant temperature approximation $\rho(E) = \rho_0 e^{E/T_{LD}}$, where $T_{LD}$ is called simply $T$ in the figures.

The level density $\rho(E)$ is another quantity which characterizes the ensemble of chaotic states by the thermodynamic temperature $T_{LD}$. It is discussed in Refs. [5,6], which review the extended earlier work. Interpreting

$$S(E) = \ln[\rho(E)], \quad \frac{1}{T_{LD}} = \frac{dS(E)}{dE} \tag{3}$$

as the microcanonical entropy, $T_{LD}$ is given by the standard relation of thermodynamics. It is a key factor in determining the spectral distribution $P(E_\gamma)$ of the emitted radiation. Fig. 5 demonstrates that for [60,68]Fe below an excitation energy of 5 MeV $\rho(E)$ can be approximated by a constant temperature. This is the case for the other example and a well know experimental fact For most examples discussed, the parameter $T_B$ of LEMAR differs from $T_{LD}$. The relation between $T_B$ and $T_{LD}$ remains open at this point.

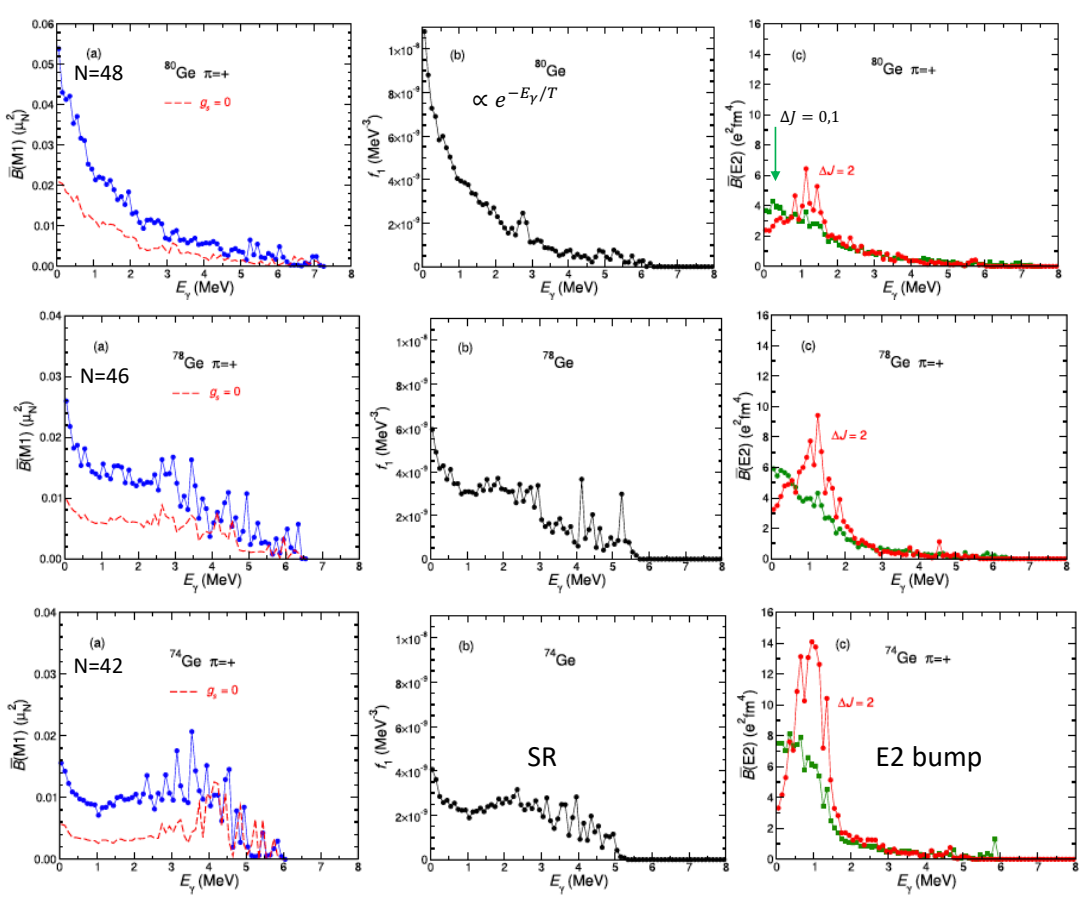

**Fig. 6.** The middle column displays the emission $\gamma sf$ of the Ge isotopes calculated by means of the SSM in Ref. [5]. The red full curves in the right column show the average $\overline{B(E2, I \to I-2)}$ values and the green curves the average $\overline{B(E2, I \to I-1)}$ values. The left column shows the average $\overline{B(M1)}$ for emission, where the red dashed curves the show the orbital part of $\overline{B(M1)}$ only. From Ref. [7].

## 3 The Ge isotopes

The discussion implies that, quite generally, one expects for the M1 emission $\gamma sf$ a single LEMAR spike at $E_\gamma = 0$ for nuclei with a small deformation and a bimodal profile of LEMAR at $E_\gamma = 0$ and a SR at $E_\gamma \sim 3$ MeV. The study of the Ge isotope chain in Ref. [7] confirms this. The authors used the SSM with a suitable configuration space and Hamiltonian. Figs. 6 and 7 show the emission $\gamma sf$. The figures also display the average values of the $B(E2, I \rightarrow I-2)$ transitions, which are a measure of the deformation of the excited states.
Traversing the neutron shell from the top, *N*=48 has a single LEMAR spike with a shoulder and a small $B(E2, I \rightarrow I-2)$ bump. With decreasing *N*, the SR appears in addition to the LEMAR spike while the $\mathrm{B}(E2, I \rightarrow I-2)$ bump grows. The trend saturates at *N*=38. As expected, the $B(E2, I \rightarrow I-2)$ bump decreases for *N*=34, 32 toward the shell bottom. However, instead of the expected single LEMAR spike a low flat region develops. The reason is that the large part of the M1 operator changes isospin, which strongly suppresses the low energy M1 transitions in $N \approx Z$ nuclei

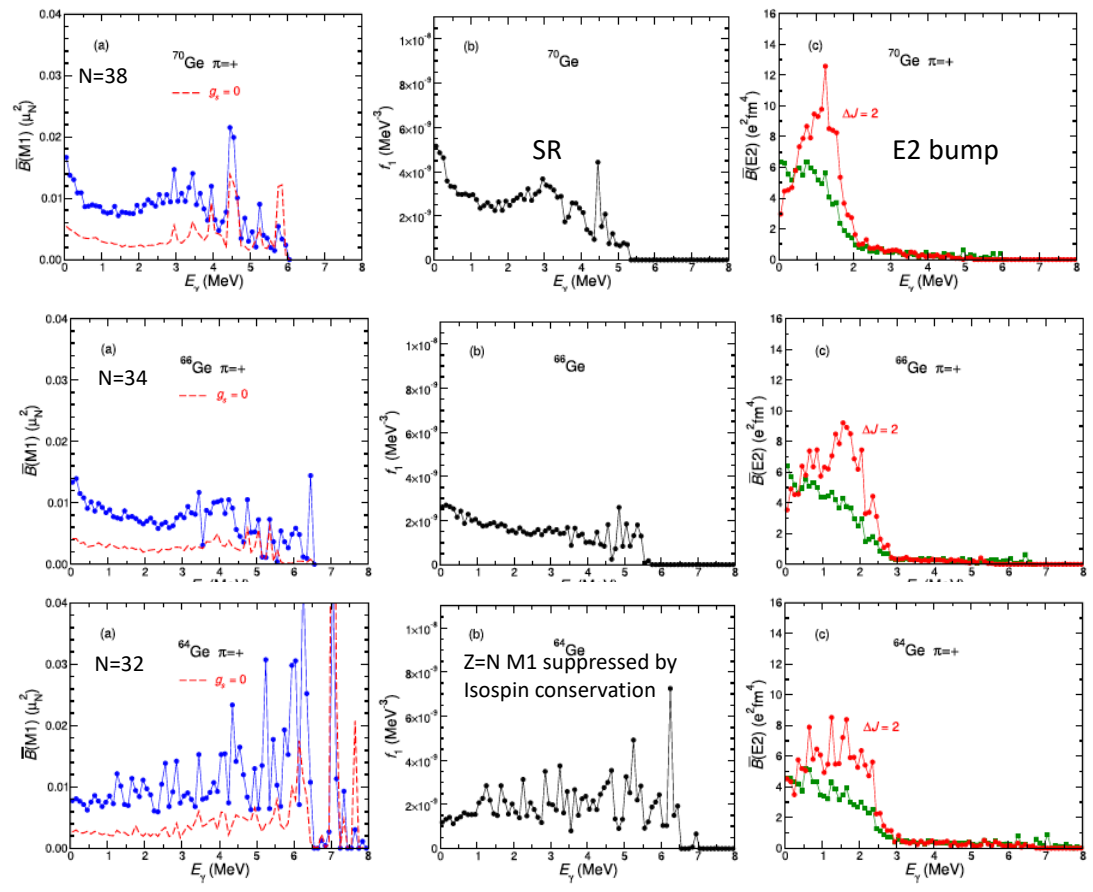

**Fig. 7.** Fig. 6 continued. From Ref. [7].

TABLE II. Summed $B(M1)$ strengths in ranges of transition energy (LEMAR, $E_\gamma < 2$ MeV; scissors, 2 MeV $\leqslant E_\gamma < 5$ MeV; $\sum$, the sum of the two), summed $B(E2)$ strengths for $E_\gamma < 5$ MeV, and summed strengths of all discrete transitions from $1^+$ states to the ground states in $^{64,66,70,74,78,80}$Ge.

| | $B(M1)_{\mathrm{tot}}$[a] ($\mu_N^2$) | | | $B(E2)_{\mathrm{tot}}$[b] ($e^2$ fm$^4$) | $\sum B(M1, 1^+ \rightarrow 0_1^+)$[c] ($\mu_N^2$) | |
|---|---|---|---|---|---|---|
| | LEMAR | SR | $\sum$ | | EXP[d] | CALC |
| $^{64}$Ge$_{32}$ | 0.30 | 0.54 | 0.84 | 155 | | 0.001 |
| $^{66}$Ge$_{34}$ | 0.35 | 0.35 | 0.70 | 185 | | 0.25 |
| $^{70}$Ge$_{38}$ | 0.54 | 0.62 | 1.16 | 219 | 0.04(1) | 0.49 |
| $^{74}$Ge$_{42}$ | 0.44 | 0.50 | 0.94 | 241 | 0.30(3) | 0.57 |
| $^{78}$Ge$_{46}$ | 0.63 | 0.49 | 1.12 | 181 | | 0.48 |
| $^{80}$Ge$_{48}$ | 0.84 | 0.28 | 1.12 | 123 | | 0.31 |

[a]Integrated $M1$ strength calculated according to Eq. (2).
[b]Integrated $E2$ strength calculated for positive-parity states in analogy to Eqs. (1) and (2).
[c]Summed $M1$ strength of transitions from the $1^+$ states below 5 MeV to the ground state.

**Tab. 1.** The summed emission $B(M1)_{tot}$, $B(E2, I \rightarrow I-2)_{tot}$ and absorption $B(M1, 0^+ \rightarrow 1^+)/3$ for the Ge isotopes. From Ref. [7].

Tab. 1 demonstrates the redistribution of the summed $B(M1)_{sum}$ strength from the LEMAR region $0 < E_\gamma < 2$ MeV to the SR region 2 MeV $< E_\gamma < 5$ MeV. The sum of the two regions decreases toward the shell bottom, which reflects the M1 suppression near $N = Z$. After removing the statistical weight of 3, the sum of the B(M1) of the absorption lines up to 5 MeV is only about one half of the sum of the emission lines.

## 4 The Sn isotopes

The authors of Ref. [8] investigated the M1 $\gamma sf$ in the Sn isotopes in the framework of the SSM using a configuration space and Hamiltonian that were well tested for the lower part of the neutron shell. Figs. 8, 9 traverse the neutron shell of these semi magic nuclides from the top. The $\gamma sf$ changes with *N* in the same way as for the above discussed examples with protons in the open shell. For *N*=80 there is only the LEMAR spike seen in the $\gamma sf$. For *N*=74 a small SR peak near 4 MeV appears in the emission $\gamma sf$ in addition to the LEMAR spike. Weak SR absorption lines are seen around 5 MeV. For *N*=68 a strong SR develops in the emission $\gamma sf$ jn addition to the LEMAR spike. A SR group of lines at somewhat higher energy is seen in the absorption spectrum. The average $\overline{B(E2)}$ values in the region $E_\gamma < 2$ MeV increase with *N*= 80, 74, 68.

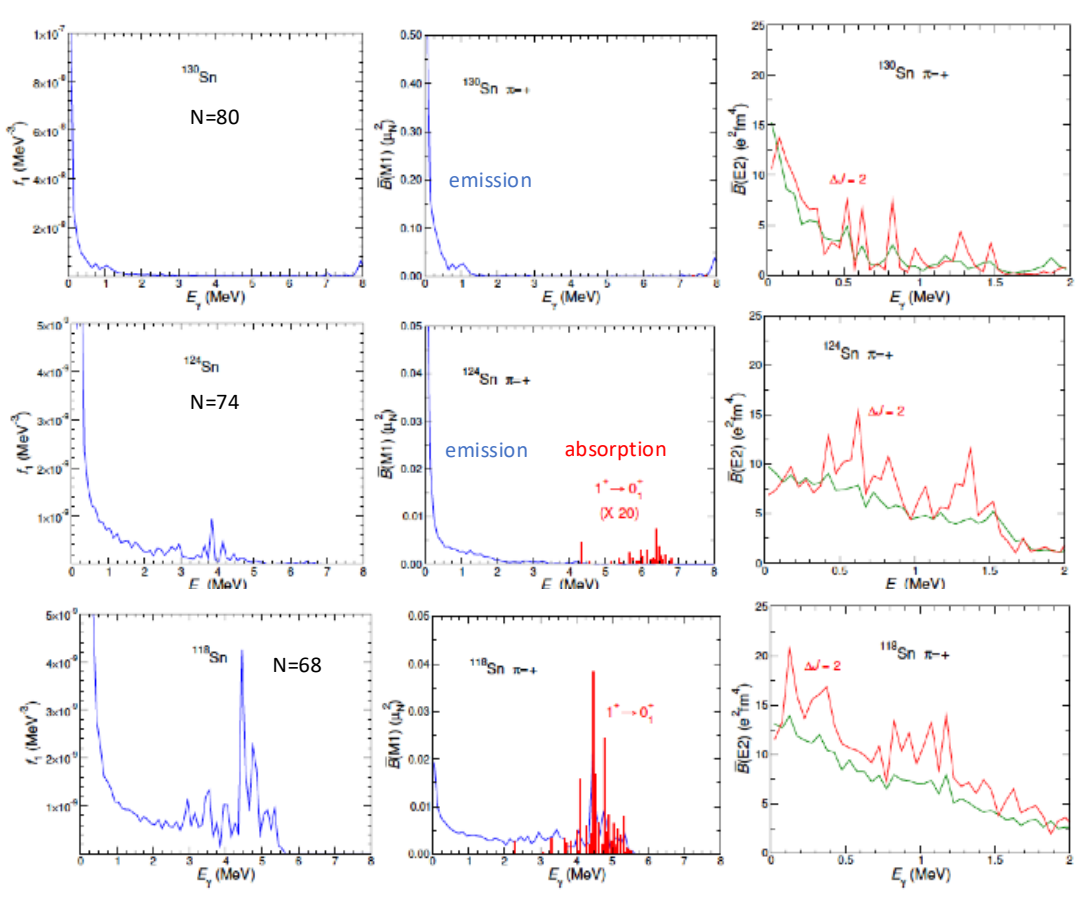

**Fig. 8.** The left column shows the emission $\gamma sf$ of the Sn isotopes calculated by means of the SSM in Ref. [6]. The middle column displays the average $\overline{B(M1)}$ for emission and the absorption lines from the ground state in Red. In the right column the green full curves show the average $\overline{B(E2, I \rightarrow I-2)}$values and the red curves the average $\overline{B(E2, I \rightarrow I-1)}$ values. From Ref. [8].

The bimodal structure persists for *N*=60, 58, 56, which is consistent with large average $\overline{B(E2)}$ values. The LEMAR parameter for all isotopes, except *N*=80, is $T_B \approx 0.3$ MeV, which is about one half of the thermodynamic temperature $T_{LD} \approx 0.6$ MeV.
Tab. 2 lists the summed emission $B(M1)_{tot}$ for the LEMAR and SR regions. A substantial SR part of $B(M1)_{tot}$ coexists with the LEMAR part for *N*= 68, 62, 60, 58, 56. The SR part is small for *N*=80, 74, 62, which is associated with closing the $h_{11/2}$ and $g_{7/2}$ shells, respectively. The suppression of the SR in the absorption spectrum is seen at the same neutron number. The large LEMAR value for $N$=80 should be disregarded as an artifact caused by the too small configuration of only two neutron holes.

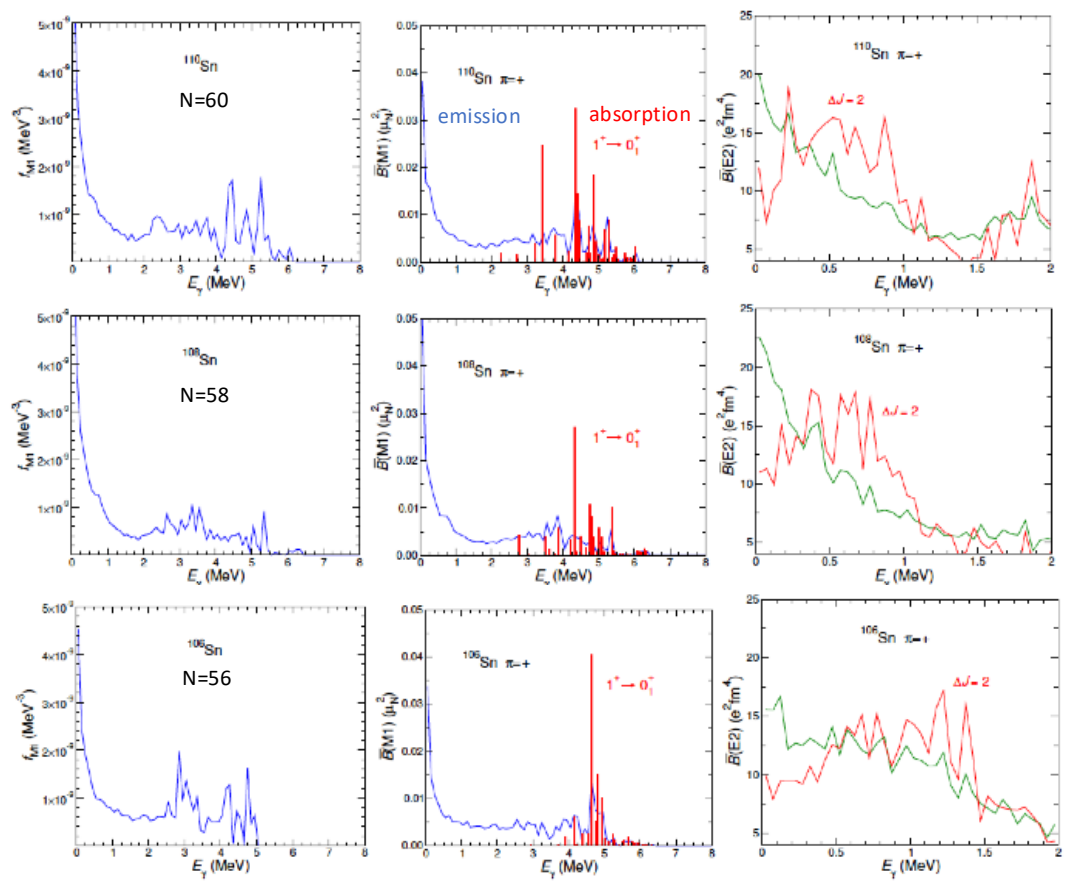

**Fig. 9.** Fig. 8 continued. From. Ref. [8].

TABLE II: Summed $B(M1)$ strengths in ranges of transition energy (LEMAR region: $E_\gamma < 2$ MeV, Scissors region: 2 MeV $\leq E_\gamma < 5$ MeV, the sum of the two) and summed $M1$ strengths from $1^+$ states up to 5 MeV to the ground state.

| | $B(M1)_{\rm sum}$ $(\mu_N^2)$ [a] | | | |
|---|---|---|---|---|
| | LEMAR | Scissors | Both | $1^+ \to 0_1^+$ [b] |
| $^{106}$Sn$_{56}$ | 0.194 | 0.190 | 0.384 | 0.085 |
| $^{108}$Sn$_{58}$ | 0.236 | 0.131 | 0.367 | 0.076 |
| $^{110}$Sn$_{60}$ | 0.223 | 0.186 | 0.409 | 0.131 |
| $^{112}$Sn$_{62}$ | 0.486 | 0.055 | 0.541 | 0.001 |
| $^{118}$Sn$_{68}$ | 0.502 | 0.239 | 0.741 | 0.155 |
| $^{124}$Sn$_{74}$ | 0.349 | 0.056 | 0.405 | 0.0003 |
| $^{130}$Sn$_{80}$ | 1.981 | 0.047 | 2.028 | 0 |

[a]Summed $M1$ strength in a given energy range calculated according to $B(M1)_{\rm sum} = 9/(16\pi)\ (\hbar c)^3\ \sum f_{M1}(E_\gamma)\Delta E_\gamma$.
[b]Summed $M1$ strength of transitions from the $1^+$ states below 5 MeV to the ground state.

**Tab. 2.** The summed M1 emission $\gamma sf$ for the Sn isotopes and the sum of the summed absorption $B(M1, 0^+ \to 1^+)/3$ for the Sn isotopes. From Ref. [8].

## 5 The Gd isotopes

The authors of Ref. [9] studied the Gd isotopes by means of the Triaxial Projected Shell Model (TPSM) [10]. The model spans the configuration space by projecting states of good angular momentum from triaxial quasiparticle configurations, where the deformation parameters $\varepsilon$ and $\gamma$ of the triaxial potential mean field are input parameters of the TPSM calculation. A Pairing+Quadrupole-Quadrupole (P+QQ) Hamiltonian was diagonalized, the coupling constants of which were fixed by requiring selfconsistency with the deformation and the pair gaps of the mean field. The TPSM calculations were carried out including two-quasi proton, two-quasi-neutron and two-quasi proton-two-quasi-neutron configurations. As for the SSM, the $B(M1, I \to I, I \pm 1)$ were calculated and sorted to obtain the $\gamma sf$.

Fig. 10 displays the $\gamma sf$ of the Gd isotopes. The weakly deformed isotopes with $N$=84, 88, have a single LEMAR spike, where the deformed ones with $N$=90 - 100 show the bimodal LEMAR-SR profile with the SR centred at 2.5 MeV. The appearance of the bump at 2.5 MeV for $N$=86 does not fit the trend. The reason is not clear.

The TPSM predicts only one strong absorption line at $E_\gamma = 2.5 - 2.8$ MeV. Other lines are tiny. In experiment one observes several absorption lines of comparable M1 strength distributed around 3.2 MeV over few 100 keV.

The LEMAR is characterized a small parameeter $T_B =$ 0.22 MeV for $N$<96 and 0.13 MeV for $N \leq 96$ and large thermodynamic temperatures of $T_{LD}$ =1.0 MeV. This suggests that the P+QQ interaction causes a weaker randomization of the high lying states than the interactions the SSM diagonalizations.

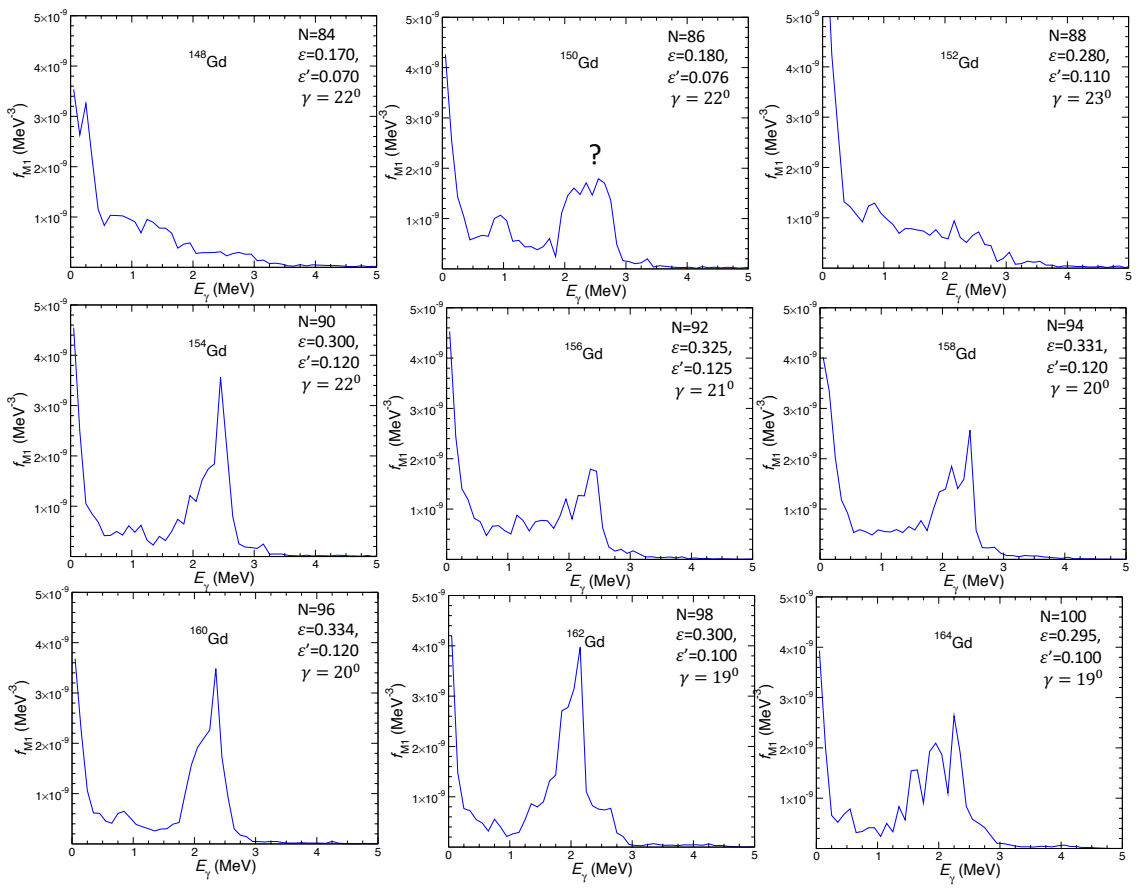

**Fig. 10.** The emission $\gamma sf$ of the Gd isotopes calculated by means of the triaxial projected shell model in Ref. [9]. From Ref. [9].

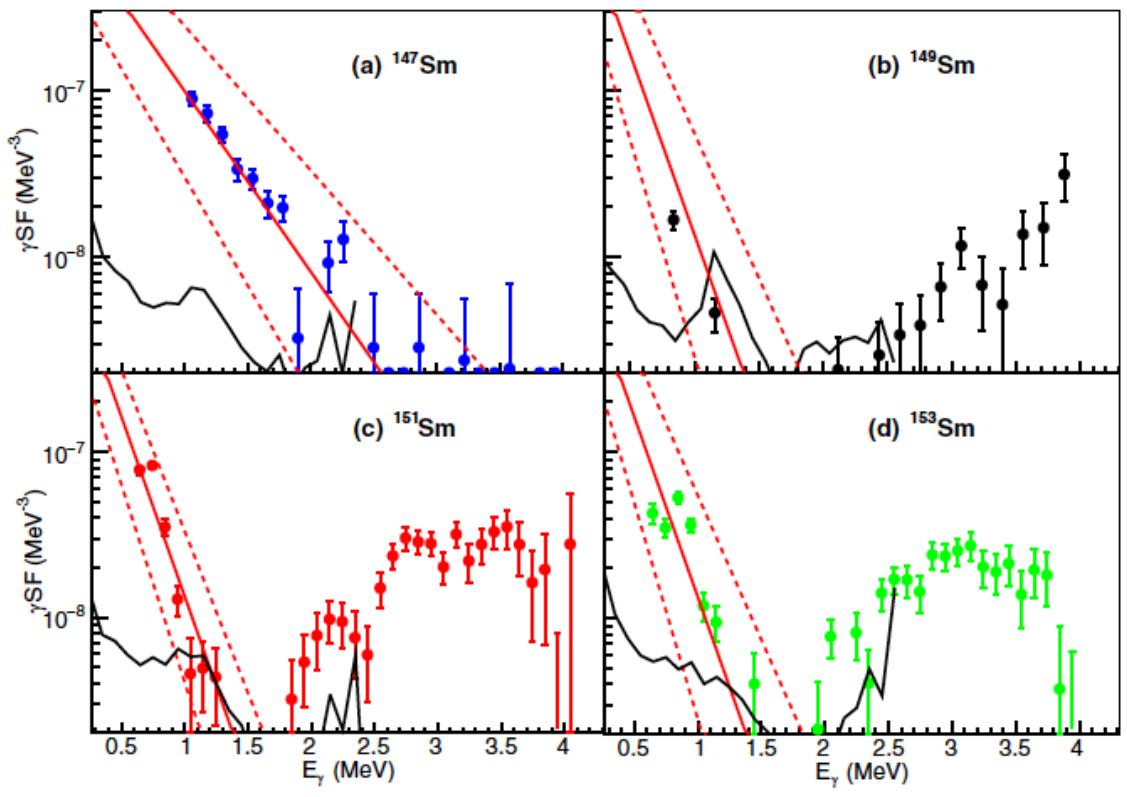

**Fig. 11.** The experimental emission $\gamma sf$ of $^{147,149,151,153}$Sm$_{85,87,89,91}$. The straight lines display a constant $T_B$ fit to the LEMAR spike and the estimated uncertainties. The black lines show the results of a SSM calculation. From Ref. [11].

## 6 The Sm isotopes and experiment

The study of the emission $\gamma sf$ of the Sm isotopes in Ref. [11] is the only clear evidence that supports the transition from the monomodal LEMAR profile to the bimodal LEMAR+SR profile with increasing deformation. As seen in Fig 11, the transition to a deformed shape occurs at $N$=88, which is well known from properties the low-lying E2 excitations

The LEMAR parameter is estimated as

$T_B = 0.3, 0.2, 0.2, 0.2$ MeV for $N$=85, 87, 89, 91, respectively where the estimate of thermodynamical temperature $T_{LD}$ =0.5 MeV is the same for all $N$.
An SSM calculation was carried out in a highly truncated configuration space. Qualitatively it reproduces the transition from the monomodal to the bimodal profile, however it underestimates the total strength by a factor of 50.
Most of the examples of the emission $\gamma sf$ do not allow to estimate $T_B$ with any satisfactory accuracy because reliable data on $\overleftarrow{f}_1(E_\gamma)$ below 2 MeV are missing. Exceptions are the discussed Sm isotopes, $^{94,96,97}$Mo [12], for which one finds $T_B$=0.3 MeV and $T_{LD}$=0.7 MeV, and $^{96,97}$Fe [13], for which one finds $T_B$=0.4 MeV and $T_{LD}$ ≈1.2 MeV.
Tab. 3 lists $T_B$ and $T_{LD}$ extracted from the experimental and theoretical results considered in this talk. For most cases the ratios $T_B/LD$~1:2 - 1:3, except the ratios ~1:1 for the Ge isotopes, where the quoted $T_B$ values are rather uncertain. The large ratio for the Gd isotopes signalizes a weak randomization of the wave functions by the P+QQ interaction.

| Nucleus | $T_B$(MeV) | $T_{LD}$(MeV) | source |
|---|---|---|---|
| $^{94,96,97}$Mo | 0.3 | 0.7 | exp |
| $^{94}$Mo | 0.4 | 0.7 | SSM |
| $^{56,57}$Fe | 0.4 | 1.4, 1.0 | exp |
| $^{60,,64}$Fe | 0.4 | 1.0, 1.2 | SSM |
| $^{70\text{-}80}$Ge | 1.0 | 0.8 | SSM |
| $^{106\text{-}124}$Sn | 0.3 | 0.6 | SSM |
| $^{147\text{-}153}$Sm | 0.3, 0.2 | 0.5 | exp |
| $^{148\text{-}164}$Gd | 0.2, 0.1 | 1.0 | TPSM |

**Tab. 3.** LEMAR parameters and thermodynamic temperatures of the discussed examples.

## 7 Summary

New SSM and TPSM calculations of the emission $\gamma sf$ for the Ge, Sn and Gd isotopes confirm the transition from the mono modal LEMAR spike in weakly deformed nuclei to the bimodal LEMAR+SR profile in deformed nuclei with a redistribution of the M1 strength, which was found before for the Mo and Fe isotopes. This dependence on valence particle number is expected to appear quite generally because it is caused by the splitting of j-shell orbitals by deformation and the randomizing of the wave functions by the residual interaction, which are robust features. The LEMAR part of the $\gamma sf$ falls off exponentially with the transition energy near $E_\gamma = 0$, which suggests a stochastic character. The associated parameter $T_B$ may be a new measure for the degree of chaoticity of the compound states. It differs from the thermodynamical temperature, derived from the level density, which is larger for most of the examples. The interpretation of their relation is an interesting open problem.

## References

1. G. A. Bartholomew, E. Earle, A. J. Ferguson, J. W. Knowles, and M. A. Lone, in Gamma-Ray Strength Functions, edited by M. Baranger and E. Vogt, Advances in Nuclear Physic (Springer, Boston, 1973), p. 229.
2. M. Guttormsen, R. Chankova, U. Agvaanluvsan, E. Algin, L. A. Bernstein, F. Ingebretsen, T. Lönroth, S. Messelt, G. E. Mitchell, J. Rekstad, A. Schiller, S. Siem, A. C. Sunde, A. Voinov, and S. Ødegaard, Radiative strength functions in 93-98 Mo, Phys. Rev. C 71, 044307 (2005). https://doi.org/10.1103/PhysRevC.71.044307
3. R. Schwengner, S. Frauendorf, and A. C. Larsen, Phys. Rev. Lett. 111, 232504 (2013). https://doi.org/10.1103/PhysRevLett.111.232504
4. R. Schwengner, S. Frauendorf, and B. A. Brown, Low-Energy Magnetic Dipole Radiation in Open-Shell Nuclei, Phys. Rev. Lett. 118, 092502 (2017). https://doi.org/10.1103/PhysRevLett.118.092502
5. *Aage* Bohr and *Ben* Mottelson, Nuclear Structure I, (W. A. Benjamin, Inc. 1969), p. 302 ff.
6. V. Zelevinsky, Quantum Chaos and Complexity in Nuclei, Ann. Rev. Nucl. Part. Sci. 46:237-79 (1996) https://doi.org/101146/annurev.nucl.46.1.327
7. S.Frauendorf, R. Schwengner, Evolution of low-lying M1 modes in germanium isotopes, Phys. Rev. C 105, 034335 (2022). https://doi.org/0.1103/PhysRevC.105.034335
8. S. Frauendorf and R. Schwengner, Low-Energy Magnetic Dipole Radiation in the Tin Isotopes, Phys. Rev. C (to be published).
9. G. Bhat, S. Frauendorf, R. Schwengner, J. Sheikh, Low-Energy Magnetic Dipole Radiation in the Gadolinium Isotopes, Phys. Rev. C (to be published).
10. J. A. Sheikh and K. Hara, Projected Triaxial Shell Model Approach, Phys. Rev. Lett. 82, 3968 (1999)
11. F. Naqvi, A. Simon, M. Guttormsen, R. Schwengner, S. Frauendorf, C. S. Reingold, J. T. Burke, N. Cooper, R. O. Hughes, S. Ota and A. Saastamoinen, Nuclear level densities and γ-ray strength functions in samarium isotopes, Phys. Rev. C 99, 054331 (2019) https://doi.org/10.1103/PhysRevC.99.054331
12. M. Guttormsen R. Chankova, U. Agvaanluvsan, E. Algin, L. A. Bernstein, F. Ingebretsen, T. Lönnroth, S. Messelt, G. E. Mitchell, J. Rekstad, A. Schiller, S. Siem, A. C. Sunde, A. Voinov, and S. Ødega˚rd, Radiative strength functions in $^{93-98}$Mo, Phys. Rev. C 71, 044307 (2005) https://doi.org/10.1103/PhysRevC.71.044307
13. A. C. Larsen, M. Guttormsen, N. Blasi, A. Bracco, F. Camera, L. Crespo Campo, T. K. Eriksen, A. Görgen, T. W. Hagen, V. W. Ingeberg, B. V. Kheswa, S. Leoni, J. E. Midtbø, B. Million, H. T. Nyhus, T. Renstrøm, S. J. Rose, I. E. Ruud, S. Siem, T. G. Tornyi, G. M. Tveten, A. V. Voinov, M. Wiedeking and F. Zeiser, Low-energy enhancement and fluctuations of γ-ray strength functions in $^{56,57}$Fe: test of the Brink–Axel hypothesis, J. Phys. G: Nucl. Part. Phys. 44 (2017) 064005, https://doi.org/10.1088/1361-6471/aa644a